\documentclass[a4paper,twoside]{article}

\usepackage{epsfig}
\usepackage{subcaption}
\usepackage{calc}
\usepackage{amssymb}
\usepackage{amstext}
\usepackage{amsmath}
\usepackage{amsthm}
\usepackage{multicol}
\usepackage{pslatex}
\usepackage[bottom]{footmisc}
\usepackage{SCITEPRESS}     

\usepackage{graphicx} 

\begin{document}

\title{Application of property-based testing tools\\ for metamorphic testing}

\author{\authorname{Nasser Alzahrani, Maria Spichkova, James Harland}
\affiliation{School of Computing Technologies, RMIT University, Melbourne, Australia} 
\email{s3297335@student.rmit.edu.au, \{maria.spichkova,james.harland\}@rmit.edu.au}
}

\abstract{
Metamorphic testing (MT) is a general approach for the testing of a specific kind of software systems -- so-called ``non-testable'', where the ``classical'' testing approaches are difficult to apply. 
MT is an effective approach for addressing the test oracle problem and test case generation problem. The test oracle problem is when it is difficult to determine the correct expected output of a particular test case or to determine whether the actual outputs agree with the expected outcomes. The core concept in MT is metamorphic relations (MRs) which provide formal specification of the system under test.  One of the challenges in MT is \emph{effective test generation}. Property-based testing (PBT) is a testing methodology in which test cases are generated according to desired properties of the software. In some sense, MT can be seen as a very specific kind of PBT.\\
In this paper, we show how to use PBT tools to automate test generation and verification of MT. In addition to automation benefit, the proposed method shows how to combine general PBT with MT under the same testing framework.
\\
~\\
\emph{Preprint. Accepted to the 17th International Conference on Evaluation of Novel Approaches to Software
Engineering (ENASE 2022). Final version published by SCITEPRESS, http://www.scitepress.org}
}

\onecolumn \maketitle \normalsize \setcounter{footnote}{0} \vfill

\section{\uppercase{Introduction}}
\label{introduction}

Formal specification is an essential tool for managing the complexity of specifying and verifying the design and the development of critical software systems. The formal approach removes ambiguity, improves precision, and used to verify that the requirements are fulfilled. 
Appel et al. summarised a number of desired qualities that the specification should have in order to be effective, see
\cite{appel2017position}.  Firstly, the specification has to be \emph{formal}, where the specification should be mathematically precise. It should be \emph{rich}, i.e. precisely expressing the intended behaviour of the system (we could reformulate this quality as \emph{completeness}). The specification has to be \emph{two-sided} where the specification is exercised by both implementations and clients. Finally, the specification has to be \emph{live} where it is automatically checked against actual code rather than some abstract model.

Formal languages like TLA+ \cite{lamport2002specifying}, 
or Alloy \cite{jackson2012software} are generally concerned with specifying systems against some models rather than the actual code under development. On the other hand, property-based testing (PBT) facilitates the use of formal specifications on actual code which help in finding subtle faults on live running systems \cite{hughes2016mysteries}.   
One of the main attributes of PBT is that it can automatically generate tests to cover edge cases that are not so obvious to identify manually. Two main elements of PBT approach make this possible: (1) a \emph{random test generator}, responsible for generating random values in a controlled way, and (2) a so-called \emph{shrinker}, minimizing the number of the generated tests cases to allow for easier debugging.

Metamorphic Testing (MT) is a special PBT technique elaborated for the cases where it's complicated to specify ``classical'' test cases having input and output data flows - in some cases, it's difficult to identify what could be the correct output for each particular input.
For ``classical''  testing we need to have a so-called an \emph{oracle} that can determine whether or not the
output is correct wrt. the provided input and this decision should be taken in a reasonable amount of time, see 
\cite{weyuker1982testing}. 
In the case an oracle cannot be created, the system are typically called \emph{non-testable} (or \emph{untestable}), but MT can provide an effective solution to the oracle problem using Metamorphic Relations (MR), see e.g., \cite{Segura20}.   
 MT was initially introduced in \cite{chan1998application} in the domain of numerical analysis. Since then, it has developed and been applied in many application areas, such as compilers, medical systems, embedded applications, search engines, service computing, simulation software, image processing systems, machine learning software and optimizing software 
 \cite{chen2015metamorphic}.

PBT tools for testing functional programs were first introduced in the \emph{Haskell programming language} by
\cite{claessenQuickCheckLightweightTool2000}, where 
\emph{QuickCheck}, a library for random testing of program properties, was implemented. Since then, many libraries have been developed
following this approach for different programming languages. The main components of \emph{QuickCheck} are the generator of random values, the shrinker, and the checker which runs these random values with pre-selected functions. In our work, we extend the generator and the shrinker in order to automate MT test generation and verification. 

The idea of generating random tests is not new, see for example \cite{chen2010adaptive}.  
However, using PBT tools such \emph{QuickCheck} has many advantages. First of all, PBT has many tools for automating and controlling the generation of random test cases. Secondly, these tools allow controlled strategies for generating random data of complicated data types, i.e.,  it is possible to configure how the random values distribute over the input domain. 
Although \cite{chenMetamorphicTestingReview2018} argues that one of MT's main advantages is the ease of test case automation, the MT automation is a complex task, when using PBT tools such as QuickCheck as a tool to test  MR relations.

\emph{Contributions:}
The main contribution of this paper is a systematic approach in which we utilize the random generation of test cases and automatic testing capabilities of PBT tools to automate some steps of metamorphic testing: More precisely, we automate test case generation and test case verification by extending QuickCheck's shrinker and generator with our customized shrinker and generator. These variants are simpler and more amendable to customization in the context of MT.  
 
\section{\uppercase{Background: Property-based testing}}
\label{sec:pbt}

Property-based testing (PBT) is an approach to testing software by defining general specifications and properties that must hold for all the executions of randomly generated test cases. The inputs to these test cases are random. If these properties do not hold, a minimized failing tests are reported. In PBT, test cases are generated randomly according to universally quantified properties. Examples of quantified properties include; validity checks, postconditions, model-based properties, inductive properties, and metamorphic testing.

In validity checks, one writes functions to check some invariants of the system under test or the datatypes used in the system. This process also includes writing properties to check test-case generator and test-case shrinker both produce valid results. The last step is to write property for the functions under test which performs a single random call  and checks that the return value is valid. For instance, when testing the functionality of inserting a value into a tree datatype, we demand that all the keys in a left subtree must be less than the key in the node, and all the keys in the right subtree must be greater.

A postcondition is a property that should be true after a call. One can come up with such property by asking what would be the expected state of the system after calling some function. A postcondition usually tests one function after calling this function with some random argument and then checking an expected relationship between the result and its argument. For Instance, after calling \emph{insert} (which inserts a value in some tree datatype), we should be able to find the key inserted without changing previously inserted keys. 

A model-based property is used to test some function by making a single call and comparing its result with the result of some abstract operation. The model refers to the abstraction functions which map the real arguments and results to abstract values.

Inductive properties are the properties that one can use induction to assert that the only function that can pass the tests is the correct one. This is usually done by relating a call of the function under test to calls with smaller arguments. The set of inductive properties covering all possible cases allows the testing of the base case and induction steps of an inductive proof-of-correctness.


 The canonical example in the literature to explain property based
testing is the reversing a list. One property that should hold for all
lists is that reversing a list $x$ twice returns the original list:
\[
reverse~(reverse~x) = x
\]
In the case of functional oracle-based testing, we would need to identify the equivalence partitions (the sets of inputs that have to be handled equivalently as they have to provide the same type of output but with possibly different values), and then specify at least one test case for each partition, or use boundary values in partitions that are ranges. Thus, for $reverse$ function this could be an empty list $[]$ and some non-empty list, e.g, $[1, 2, 3]$, i.e., our test cases would be
\[
\begin{array}{l}
reverse~ (reverse~[]) = []\\
reverse~ (reverse~[1,2,3]) = [1,2,3]
\end{array}
\]

\noindent 
In contrast to this approach, in PBT the checks take place on the return value of the function under test, instead of checking values ``hard-coded'' in the test cases such as $[]$ and $[1, 2, 3]$. That is, the input to the function under test would be automatically generated and the library chooses random values for testing rather than the tester specifying particular values. For example, for the $reverse$ function, one could call the $reverse$ function twice and expect the original list to be returned. The library  generates many random test cases and reports a failure if it finds a counterexample. 

In the case of \emph{QuickCheck}, the property to be passed to the library is
$reverse~(reverse~x) = x$, the library will generate the random
values for $x$, and will report a failure if it finds a counterexample.

It is worth noting that there are some anti-patterns that could emerge while writing property-based tests. Because it is sometimes difficult to think about a property, practitioners usually fall into the trap of duplicating the implementation of the code in tests. The literature on PBT has many other examples of such anti-patterns and how to avoid them. One solution to avoid this problem is MT. 

MT is a successful method in solving the oracle problem in software testing. The core idea is this: in the cases when it's hard to specify in advance what exactly should be  the output of a function, we may be able to observe the change to the output when changing the input. That is, valuable information about the function would be whether and how its input changes when we change its input. For instance, even if the expected result of a function such as inserting a key into a tree is difficult to predict, we may still be able to express an expected relationship between this result and the result of some other call. In this case, if we insert an additional key into t before calling insert k v, then we expect the additional key to appear in the result also. 

\begin{figure}[!ht]
  \centering
  \includegraphics[width=7.6cm]{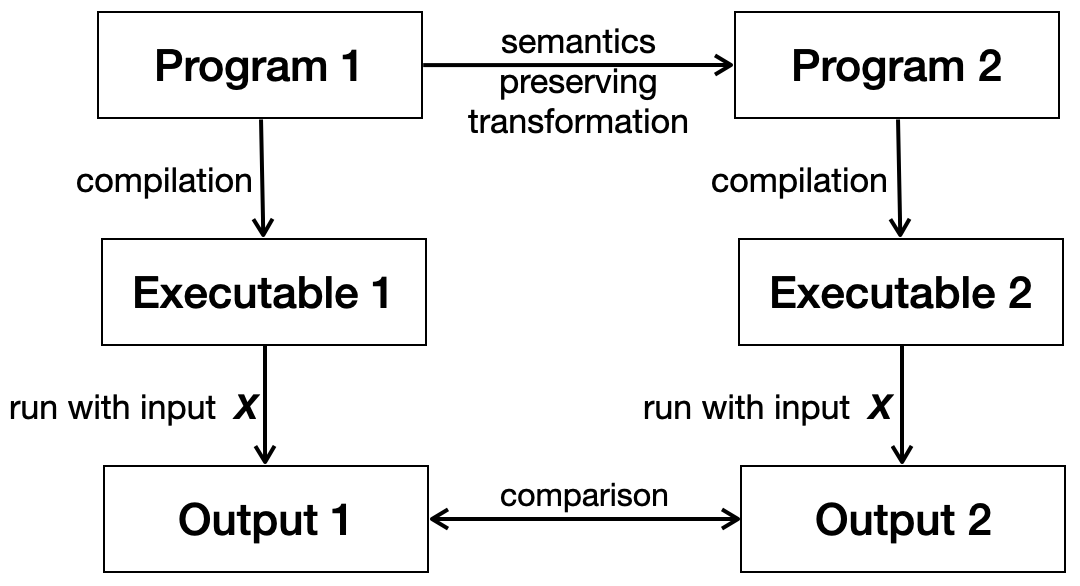}  
  \caption{Testing compilers using Metamorphic testing}
  \label{fig:mt-compiler}
\end{figure}

~\\
Metamorphic Relations (MRs) are a central element of MT. The MRs are properties of the function under test. An important (and usually missed) attribute of MR is that they relate \emph{multiple} inputs to their expected outputs. When implementing MT, we first generate source test cases. Then use the MR to generate new input. This new input is then used to compare the output from the first set of the tests with the last ones. 

In our proposed method, we show how to automate the test case generation and verification of MT using QuickCheck. Our method can be generalized to work with other PBT tools other than QuickCheck as our method extends features of QuickCheck that are common in other PBT tools.

Figure~\ref{fig:mt-compiler} illustrates MT in the context of testing a compiler. 
 There are two paths that are expected to lead to the same output:
 \begin{itemize}
     \item[(1)]
     We start with $Program~1$, it is compiled to obtain the corresponding executable code $Executable~1$. Then, we run $Executable~1$  with some input data $x$ to get an output.
     \item[(2)] 
     We modify $Program~1$ to a semantically equivalent but syntactically different $Program~2$ (e.g., by unrolling a loop or removing a comment etc.), then apply the same compilation method to get the corresponding executable code $Executable~2$. After that, we run $Executable~2$  with the same input data $x$ to get an output. 
 \end{itemize}
When both outputs are obtained, we compare them: if they are not exactly the same, the compiler is faulty. 

One of the benefits of using PBT for MT is the rich sets of tools available. For instance,  PBT tools allow the creation of strategies for generating random data for complicated data types with minimal setup. In MT, the operations and data types are usually more complicated than simple data types such as integers or lists. Existing approaches for creating these data types are ad-hoc \cite{chen2015revisit}. In these other approaches, one has to do almost the same setup work for every kind of data type in order to generate the random values. PBT tools, on the other hand, are more general and cover more data types without the need to duplicate the code for every kind of data type or model. In addition, PBT requires less code than these other approaches with more control over the distribution of the test cases space. These approaches evenly distribute the test cases over the input space. 

Creating a random BST using any PBT library requires less setup work. One only required to define the data type and pass it to the library. However, since we are using these PBT tools for MT, some more setup and customization are required. 

\section{\uppercase{Related work}}
\label{sec:related} 
The effectiveness of MT in alleviating the oracle problem has allowed it to appear in many different application domains. However, many of these applications do not provide a systematic way to automate some parts MT.

The automation of MT was first introduced in 
\cite{gotlieb2003automated} where they proposed a framework that utilizes Constraint Logic Programming techniques to find test data that violate a given metamorphic relation. However, they require the usage of special metamorphic-relations, such as permutation-based relations, to speed up the search among the possible test data. 
 
There are few other efforts to automate MT steps. For instance, Zhu 
created a tool for automating metamorphic testing for Java unit tests, see \cite{zhu2015jfuzz}. This method is specific to Java unit tests. Our method is more general and can be applied in any programming language which has library support for PBT.

  An automatic MT framework for compilers is proposed in \cite{tao2010automatic}.   
  Their approach in generating the test cases is similar to the approach presented in this paper. However, their approach is tailored to the domain of testing compilers, where we propose a generally applicable solution.

In \cite{liu2012new}, the authors propose a method that allows the composition of new  metamorphic relations based on previously defined ones, their case study showed that new metamorphic relations can be constructed by compositing some existing metamorphic relations. They assert that the new derived metamorphic relation delivers better metamorphic testing than the original metamorphic relation as well as reduces the number of test cases.
 
 Work related to verifying authenticated data structures (ADS) is presented in \cite{miller2014authenticated}. The approach of Miller provides a semantics for a programming language \emph{LambdaAuth}, which  supports ADS. This approach provides many benefits, however, 
 it might be hard to convince practitioners to use it which is less likely to be widely spread among engineers and is difficult to have an impact. In \cite{brun2019generic}, the authors used Isabelle proof assistant to formally verify \emph{LambdaAuth}. They also assert that they found several mistakes in the semantics of \emph{lambdaAuth}. In our work, we use a mainstream programming language (\emph{Haskell}) to design such ADS and verify our implementation of these ADS using PBT and MT.

 \section{\uppercase{Proposed Approach}}
\label{sec:approach}

In this section, we present a systemic method to use PBT tools to test MRs. We specifically choose \emph{QuickCheck} to illustrate the proposed approach. Our method is general and can be implemented using other PBT libraries as well. Our proposed method consists of three aspects. First, we develop a new generator for generating test cases for MT. Second, we develop a new test case shrinker. Finally, we use the newly designed generator and the shrinker instead of \emph{QuickCheck}'s default generator and shrinker.  In the rest of the section, we present the core features and steps of our approach, and then discuss the advantages of this approach.


\emph{QuickCheck} is a library for random testing of program properties. The programmer provides a specification of the program, in the form of properties that functions should satisfy. The library then generates a large number of random test cases and checks that the property holds. Specifications are expressed in Haskell. The Haskell programming language also provides functions to define properties, observe the distribution of test data, and define test data generators, which is an important advantage for system specification.

When using PBT tools such as \emph{QuickCheck} there is some expected setup that needs to be done before defining the properties. One such setup is \emph{shrinking}. The main objective of \emph{shrinking} is to produce a minimum failing test cases which facilitate the debugging of the program. Another required setup is the \emph{random values generator} which can be configured depending on the scenario. More importantly, the generator and shrinker need to be designed to work together when testing MRs. Otherwise, if we use the default test case generator and shrinker, the checkers might miss some test cases or generate invalid tests. The proposed method ensures that does not happen. 

The steps needed to systematically test MR relations using PBT tools are as follows: 

\begin{itemize}
    \item[(1)] Specify an  
    MR property
    \item[(2)] Customize the test case generator
    \item[(3)] Customize the test case shrinker
    \item[(4)] Run the checker
\end{itemize}
Our main contributions are in Steps 2 and 3. 
Let us now discuss these steps and our solution in more detail.

\textbf{Step 1: Specify an MR property.} 
Specifying suitable MRs is key in MT. Although identifying MRs is not a difficult task, this is typically a manual procedure, see \cite{mayer2006empirical}, and we don't intend to automate it within our approach.
However, there have been some approaches intending to automate this step \cite{chen2016metric}, and in our future work, we consider combining our method with these approaches.

\textbf{Step 2: Customize the test case generator.}  
The first thing that all PBT libraries do is to generate random inputs for the functions under test. In the PBT literature, this is known as \emph{generation}. For every type, there is an associated random test generator. 

For example, to generate a list of values, one has to  use the generator together with two parameters. The first parameter is the number of elements in the list. The second parameter is the $size$ which depends on the type of values being generated and the context. For example, the $size$ can be the maximum value of Integer type, the maximum length of a list, or the depth of a binary search tree.

For MT, the generator has to be customized to produce valid test cases. In Section~\ref{sec:example} we will present an example of BST, where the values should not be generated uniformly. 


\textbf{Step 3: Customize the test case shrinker.}   
Almost all PBT libraries and frameworks have a mechanism to reduce the set of generated test cases that fails a property to a minimum number of failing test cases that is necessary for the debugging process, as an unnecessary large number of the failed tests cases (where many cases might refer to the same error) will make debugging process more complicated and time-consuming. This mechanism is known in the literature as \emph{shrinking}.  

\textbf{Step 4: Run the checker.}  
In this step, we pass the MR property, which was specified in  \emph{Step 1} to the checker. If the function under test does not satisfy the MR, the PBT library will report the failing test case that violated the MR property.

~\\
As mentioned in \emph{Step 3}, we cannot use the default \emph{QuickCheck's} shrinker for testing MR properties. Thus, we modify \emph{QuickCheck} generator with our designed generator. 
The default \emph{QuickCheck}'s generator is based on  \cite{claessen2013splittable} which is found to require some efforts to use in MT. On this basis, we design a modified generator and instruct \emph{QuickCheck} to use it instead of its default one. The advantage of the proposed approach is 
that the testing of MT can together with other properties under the same testing framework. 
Thus, the same shrinker and verifier can be used for both MT and general properties to test. 

Our version of shrinker has the following features:
\begin{itemize}
    \item The values are enumerated by depth instead of size and for this reason, the number of values tends to grow quickly as our shrinker explores further test cases.
    \item The modified shrinker exploits laziness \cite{hudak2007history}. That is it uses partially defined test values. If a property returns a Boolean result for a partially defined value, the shrinker does not enumerate more versions of this value. The benefit is that the checker will stop as soon it encounters the first failing test which improves the speed of the checker.
\end{itemize}

One of the differences between QuickCheck shrinker and our shrinker is that our method of shrinking is integrated into generation. It is worth noting that almost all PBT tools in many different programming languages use a similar shrinking methodology as the one used in QuickCheck. The main problem with this approach is that  shrinking is defined based on datatypes. This constraints the ways in which values are shrinked. That is, there is only one way to define shrinking for the same data type without taking into consideration the way it was generated. On the other hand, our shrinker is composed with the generator and the generator controls how the values it produces shrinks. 

Our approach to shrinking has many benefits. For example, shrinking happens even if there is no defined shrinker on the datatype. This allows the shrinker to share the same variants as the generator and, at the same time, reduce the effort needed to write a separate shrinker for each datatype involved in the test. Another benefit of our shrinker is that failure reported is more revealing than the shrinker defined as datatype. For instance, in \emph{QuickCheck}, errors are sometimes shrinked to different errors, which is undesirable since the error we expect is being reduced to another error we do not care about. To mitigate this problem, one has to duplicate the constraint logic both in the generator and in the shrinker.  In our implemented shrinker, the main idea is that we shrink the outputs by shrinking the inputs. This help in finding possible more shrinks based on that representation. 

Our designed shrinker covers the range between the smallest value of some type and increases the value until the test fails. It repeats this process until the test passes. In this case, it reports the largest value from the previous step as the smallest test case that fails the property, i.e., the boundary values. For example, suppose that we are testing whether the value of variable $x$ of type Integer is less than 77 ($x<77$). Suppose that the first random value that is generated (by the generator) is 90 which will cause the test to fail. Then, the shrinker will generate new random values and in random steps ranging from zero to 89. Now, maybe the new failing value is 89. The shrinker will repeat the same process again for the values between zero and 88. The shrinking repeats until the random value is 78 after which the smallest failing test value is 77. After which the shrinker stops.

The way we ensure the validity of the generated (then shrinked) test cases is by adding a precondition. The main objective of a precondition is to inform the generator not to generate invalid test cases using the $valid$ function that we have to implement. The $valid$ function checks the property before passing it to the generator. The generator will still generate random test cases but they will not be executed. The $valid$ function depends on the context. For instance, in the context of Binary Search Trees, the $valid$ function checks that the keys in the left subtree are less than the key at the root node and all the keys on the right subtree is greater than the key at the root node.

\section{\uppercase{Example: Binary search tree}}
\label{sec:example}

As a running example to explain the proposed method of applying the PBT tool \emph{QuickCheck} for MT, we consider the operations of inserting into and deleting from a binary search tree (BST). This example not trivial but is simple enough to explain the proposed method. Another reason for choosing BST is that the same approach can be used for testing more elaborate kinds of trees such as \emph{Merkle} trees \cite{merkle1987digital}, see also our discussion future works in Section~\ref{sec:conclusions}. To evaluate the proposed approach, we also introduce faulty variants of the operations under test, \emph{insert} and \emph{delete}.

A BST is a type of data structure for storing values such as integers in an organized way. The internal nodes of BST store a key greater than all the keys in the node's left subtree and less than those in its right subtree. BST are usually used for fast \emph{lookup}, \emph{insert} and \emph{delete} of value items. Testing \emph{insert} function which inserts a \emph{key} and
\emph{value} in a binary search tree is difficult. Using MT approach, we can
change the input using a new \emph{key} and \emph{value} and then
observe the relationship to the original call to \emph{insert} function. MT allows more numbers of
properties to be tested. Using the example of trees, we can use
\emph{insert} with \emph{delete} and test the output. Inversely, we can
use \emph{delete} with \emph{insert} and test the output. This is true for any combination of the operations under test.

One possible mistake when testing properties of the insertion and deletion of BST, is that the test code is the same as the implementation. Therefore, if there is a bug in the implementation, it will also be in the tests which renders the tests useless. One solution to this problem is to get an appropriate metamorphic relation to test the intended behaviour. This way we can verify the correctness of the implementation without a expecting concrete output.  

\begin{figure}[!h]
  \centering 
   \includegraphics[width=7.7cm]{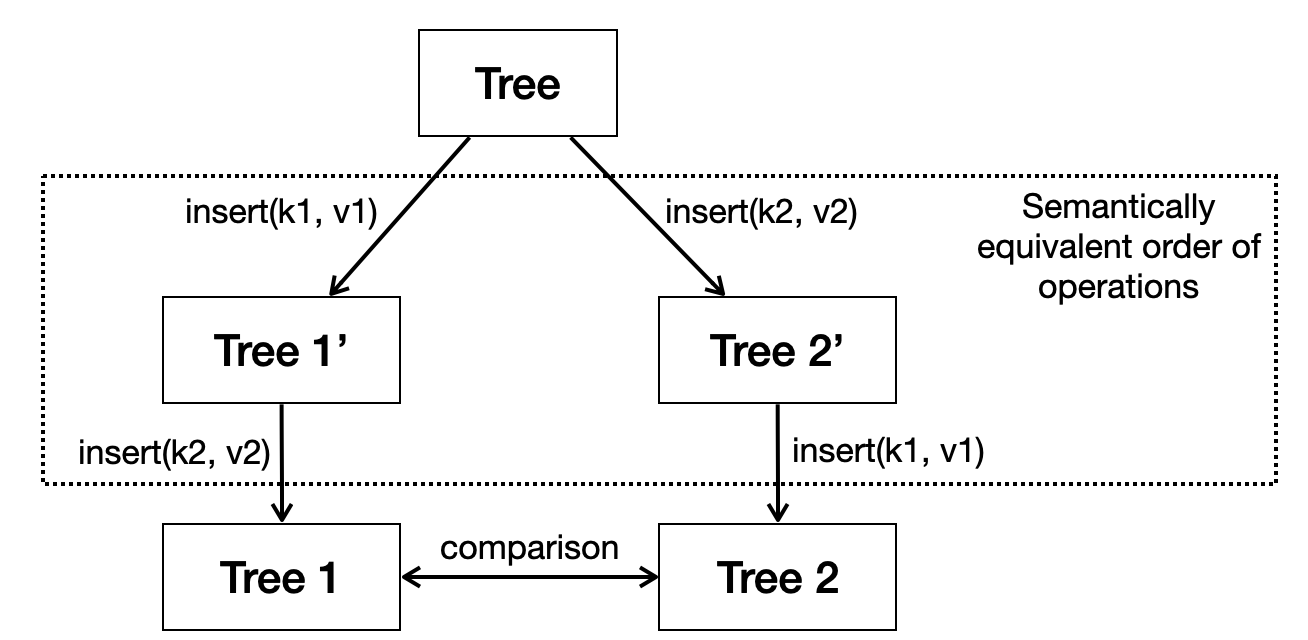} 
  \caption{MR property: \emph{Tree 1} and \emph{Tree 2} are semantically equivalent}
  \label{fig:mt-tree}
  \vspace{-0.1cm}
\end{figure}

Figure \ref{fig:mt-tree} shows the MT of inserting keys and values into a BST. Starting with the Tree at the top, we insert some key $k1$ and some value $v1$ to get some modified tree. Then, another key $k2$ and value $v2$ is inserted into the modified tree to get the $output$ tree (whatever it is).  We repeat the same operation to the original tree but we change the inputs to $insert$. That is, we insert $k2$ and $v2$ followed by inserting $k1$ and $v1$ to get the $output$ tree. The metamorphic relation asserts that the two $output$ trees should be the same otherwise $insert$ is faulty. The notion of quality between two trees depends on the operations under test. For $insert$, we can just assert that if the keys and values in both trees are the same the trees are semantically equivalent. 

~\\
To test the effectiveness of the proposed method, we intentionally introduce faulty variants of \emph{insert} and \emph{delete} and test them in a similar way. 
The faulty variants are:

\begin{itemize}
    \item[] \textbf{Fault 1} \emph{insert} removes the original tree and returns just
 the newly inserted value in a single node.
    \item[] \textbf{Fault 2} \emph{delete} does not build the tree above the key being deleted. That is, it only returns the rest of the tree instead.
\end{itemize}
 
 ~\\
Starting with the declaration of the data type, a BST for some key $k$ and value $v$, is either a $Leaf$ or a $Branch$ containing left subtree, key $k$, value $v$ and the right subtree, respectively.

\textbf{Step 1: Specify an MR property.} This is the property that we wish the PBT tool to check. Before we can choose the MR, {we need to pick the functions that we want to test}. For this example we choose \emph{insert} and \emph{delete}. $insert$ takes key $k$, value $v$ and the tree and returns the modified tree after the insertion. The $delete$ function takes key $k$ and value $v$ and returns the modified tree after the deletion.

Since we want to test two distinct functions (\emph{insert} and \emph{delete}), there are, at least, two MRs that we identify. The MRs that we want to check are the following:
\begin{itemize}
    \item \textbf{MR 1:} Inserting into the tree after modifying it with a delete operation should be the same as doing the deleting before inserting 
    \item \textbf{MR 2:} Deleting from the tree after modifying it with inserting, should be the same as doing the inserting before deleting 
\end{itemize}

Table \ref{tab:example1} shows the precise Metamorphic Relations (properties) for inserting and deleting in the context of a binary search tree. The first set shows the insertion of key $k$ and value $v$ into the tree modified by the deletion of key $k'$ from the original tree. The MR asserts this should be equivalent to deleting $k$ from the tree modified by insertion key $k$ and value $v$ into the same tree. The second is set of operations shows the deletion of key $k$ into the tree $t$ modified by the insertion of key $k'$ and value $v'$. Again, the MR asserts this should be equivalent to doing the deletion of $k$ first, then, inserting $k'$ and $v'$. 

This demonstrates how effective MT can be for generating properties. That is, if the number of operations is $n$, the number of  derived operations is $O(n^2)$, see also \cite{liu2013effectively}.

\begin{table*}[h]
\caption{Some MR properties for a BST \emph{insert} and \emph{delete}}
\label{tab:example1} 
\centering
\begin{tabular}{|c|c|c|}
  \hline
  op 1 & op 2     &    Metamorphic properties \\
  \hline
  insert & delete & \begin{tabular}[c]{@{}l@{}}
                    insert k v (delete k' t)
                    = delete k' (insert k v t)
                    \end{tabular}\\
  \hline
  delete & insert &  \begin{tabular}[c]{@{}l@{}}
                     delete k (insert k' v' t) 
                      = insert k' v' (delete k t)
                       \end{tabular} \\  
\hline
\end{tabular}
\end{table*}

\textbf{Step 2: Customize the test case generator.} We use the generator defined in section \ref{sec:approach} which generate random trees by creating a random list of keys and a random list of values and inserting them into the empty tree using $insert$ function. We also have to define $valid$ function which ensures the following:
 \begin{itemize}
     \item All the keys in the left subtree is less than the key at the root node
     \item All the keys in the right subtree is greater than the key at the root node
 \end{itemize}
   
\textbf{Step 3: Customize the test case shrinker.}
 Using the default shrink function, shrink might include invalid trees. The library may shrink the test case before reporting it. Or It may produce a valid tree with an invalid shrinking. Therefore, we must add the precondition discussed in \ref{sec:approach} to ensure only valid trees participates the shrinking process. This precondition holds for any randomly generated test. The precondition is just the $valid$ function defined in \emph{Step 2}.

\textbf{Step 4: Run the checker.} The checker is just a function that takes  any property and returns a Boolean value. We pass the MRs relations to the checker function then the library will generate many test cases. The number can be set when configuring the checker. 

For the correct variants of \emph{insert} and \emph{delete}, the PBT library reports a 100 passing tests. The number of the generated test case can also be configured to increase the assurance of the test. 
For the faulty variants, the tests report failing of tests after 100 test cases and generate the minimum failing examples for both of the introduced \emph{faults}. However, one interesting observation is that \emph{fault 1} is missed by the checker when we don't check both MR at the same time. Therefore, it is recommended to include as many MRs as necessary in a single test to specify properties of the function under test. 

One misconception of MT is that any property can be considered as an MR, see  \cite{chenMetamorphicTestingReview2018}. It is true that MR is a property but the inverse is not true. Therefore, when using PBT tools to test MR properties, we almost always use two operations, at least, in a single metamorphic test. More precisely, when using PBT tools to test metamorphic relations, we should either change the input to same function as shown in Figure \ref{fig:mt-tree} or use two distinct operations as shown in Table \ref{tab:example1}.

\section{\uppercase{Conclusions}}
\label{sec:conclusions}
In this paper, we presented a systemic method for using PBT tools to automate the test generation and verification of metamorphic relations. 
Many existing efforts for automating MT are domain-specific, i.e. the automation of the MT steps is elaborated to work only for specific application domains such as web services and specific programming languages.
The work presented in this paper is  more general and can be used in many different scenarios where MT is needed. Its advantage is in using authenticated data structures (ADS) to solve the issue.

PBT tools are  generally  used for testing universal properties other than MR such as postconditions, inductive properties, and model-based properties. In this paper, we have shown a method to created a specialized test-case generator and test-case shrinker to automate some parts of MT steps. We showed that The default shrinker is not ideal for testing some kinds of MR as it is difficult to compose previously defined MR to create new MRs. 
In addition, the default shrinker may report confusing failure cases since it is based on defining \emph{shrinking} on datatypes which forces the user to add additional duplicated code. 
However, this workaround is not needed with our shrinker since it does not have to be defined on the datatypes and there would no need to encode the invariants into the shrinkers, which requires more effort and could be difficult if the scenario is more complicated. We have implemented our method using one particular PBT tool \emph{QuickCheck}. However, our method is general and can be implemented using any other PBT tool.  

We presented our method using the Binary search tree example. The two operations we selected were \emph{insert} and \emph{delete} and we introduced faulty versions of these two operations. We showed it is recommended to use as many MRs as necessary to specify the operations under test otherwise the test might miss some subtle faults.
 
For future work, we plan to  combine the proposed approach to our earlier works presented in \cite{alzahrani2016spatio,alzahrani2017temporal},  where we used PBT tools to test models generated by formal methods tools such as TLA+ \cite{lamport1994temporal}. Another direction of the future work is to consider the human-centered aspects to enhance the proposed framework, following the ideas presented in \cite{spichkova2015human,spichkova2016human,spichkova2016ahr,spichkova2017human,zamansky2016formal}.

\bibliographystyle{plain}

\begin{thebibliography}{10}

\bibitem{alzahrani2016spatio}
Nasser Alzahrani, Maria Spichkova, and Jan~Olaf Blech.
\newblock Spatio-temporal models for formal analysis and property-based
  testing.
\newblock In {\em Federation of International Conferences on Software
  Technologies: Applications and Foundations}, pages 196--206. Springer, 2016.

\bibitem{alzahrani2017temporal}
Nasser Alzahrani, Maria Spichkova, and Jan~Olaf Blech.
\newblock From temporal models to property-based testing.
\newblock In {\em Evaluation of Novel Approaches to Software Engineering},
  pages 241--246. SciTePress, 2017.

\bibitem{appel2017position}
Andrew~W Appel, Lennart Beringer, Adam Chlipala, Benjamin~C Pierce, Zhong Shao,
  Stephanie Weirich, and Steve Zdancewic.
\newblock Position paper: the science of deep specification.
\newblock {\em Philos. Trans. R. Soc. A.}, 375(2104), 2017.

\bibitem{brun2019generic}
Matthias Brun and Dmitriy Traytel.
\newblock Generic authenticated data structures, formally.
\newblock In {\em Interactive Theorem Proving}, 2019.

\bibitem{chan1998application}
FT~Chan, TY~Chen, Shing~Chi Cheung, MF~Lau, and SM~Yiu.
\newblock Application of metamorphic testing in numerical analysis.
\newblock In {\em Int. Conf. on Software Engineering}, 1998.

\bibitem{chen2015metamorphic}
Tsong~Yueh Chen.
\newblock Metamorphic testing: A simple method for alleviating the test oracle
  problem.
\newblock In {\em Automation of Software Test}, pages 53--54. IEEE, 2015.

\bibitem{chenMetamorphicTestingReview2018}
Tsong~Yueh Chen, Fei-Ching Kuo, Huai Liu, Pak-Lok Poon, Dave Towey, T.~H. Tse,
  and Zhi~Quan Zhou.
\newblock Metamorphic {{Testing}}: {{A Review}} of {{Challenges}} and
  {{Opportunities}}.
\newblock {\em ACM Computing Surveys}, 51(1):1--27, April 2018.

\bibitem{chen2010adaptive}
Tsong~Yueh Chen, Fei-Ching Kuo, Robert~G Merkel, and TH~Tse.
\newblock Adaptive random testing: The art of test case diversity.
\newblock {\em Journal of Systems and Software}, 83(1):60--66, 2010.

\bibitem{chen2015revisit}
Tsong~Yueh Chen, Fei-Ching Kuo, Dave Towey, and Zhi~Quan Zhou.
\newblock A revisit of three studies related to random testing.
\newblock {\em Science China Information Sciences}, 58(5):1--9, 2015.

\bibitem{chen2016metric}
Tsong~Yueh Chen, Pak-Lok Poon, and Xiaoyuan Xie.
\newblock Metric: Metamorphic relation identification based on the
  category-choice framework.
\newblock {\em Journal of Systems and Software}, 116:177--190, 2016.

\bibitem{claessenQuickCheckLightweightTool2000}
Koen Claessen and John Hughes.
\newblock {{QuickCheck}}: A lightweight tool for random testing of {{Haskell}}
  programs.
\newblock In {\em Functional Programming}, pages 268--279, 2000.

\bibitem{claessen2013splittable}
Koen Claessen and Micha{\l}~H Pa{\l}ka.
\newblock Splittable pseudorandom number generators using cryptographic
  hashing.
\newblock {\em ACM SIGPLAN Notices}, 48(12):47--58, 2013.

\bibitem{gotlieb2003automated}
Arnaud Gotlieb and Bernard Botella.
\newblock Automated metamorphic testing.
\newblock In {\em Computer Software and Applications Conference}, pages 34--40.
  IEEE, 2003.

\bibitem{hudak2007history}
Paul Hudak, John Hughes, Simon Peyton~Jones, and Philip Wadler.
\newblock A history of haskell: being lazy with class.
\newblock In {\em History of programming languages}, pages 12--1, 2007.

\bibitem{hughes2016mysteries}
John Hughes, Benjamin~C Pierce, Thomas Arts, and Ulf Norell.
\newblock Mysteries of dropbox: property-based testing of a distributed
  synchronization service.
\newblock In {\em Software Testing, Verification and Validation}, pages
  135--145. IEEE, 2016.

\bibitem{jackson2012software}
Daniel Jackson.
\newblock {\em Software Abstractions: logic, language, and analysis}.
\newblock MIT press, 2012.

\bibitem{lamport1994temporal}
Leslie Lamport.
\newblock The temporal logic of actions.
\newblock {\em ACM Tran. on Programming Languages and Systems}, 16(3):872--923,
  1994.

\bibitem{lamport2002specifying}
Leslie Lamport.
\newblock {\em Specifying systems}, volume 388.
\newblock Addison-Wesley Boston, 2002.

\bibitem{liu2013effectively}
Huai Liu, Fei-Ching Kuo, Dave Towey, and Tsong~Yueh Chen.
\newblock How effectively does metamorphic testing alleviate the oracle
  problem?
\newblock {\em IEEE Transactions on Software Engineering}, 40(1):4--22, 2013.

\bibitem{liu2012new}
Huai Liu, Xuan Liu, and Tsong~Yueh Chen.
\newblock A new method for constructing metamorphic relations.
\newblock In {\em Quality Software}, pages 59--68. IEEE, 2012.

\bibitem{mayer2006empirical}
Johannes Mayer and Ralph Guderlei.
\newblock An empirical study on the selection of good metamorphic relations.
\newblock In {\em Computer Software and Applications Conference}, volume~1,
  pages 475--484. IEEE, 2006.

\bibitem{merkle1987digital}
Ralph~C Merkle.
\newblock A digital signature based on a conventional encryption function.
\newblock In {\em Theory and application of cryptographic techniques}, pages
  369--378. Springer, 1987.

\bibitem{miller2014authenticated}
Andrew Miller, Michael Hicks, Jonathan Katz, and Elaine Shi.
\newblock Authenticated data structures, generically.
\newblock {\em ACM SIGPLAN Notices}, 49(1):411--423, 2014.

\bibitem{Segura20}
S.~{Segura}, D.~{Towey}, Z.~Q. {Zhou}, and T.~Y. {Chen}.
\newblock Metamorphic testing: Testing the untestable.
\newblock {\em IEEE Software}, 37(3):46--53, 2020.

\bibitem{spichkova2015human}
Maria Spichkova, Huai Liu, Mohsen Laali, and Heinz~W Schmidt.
\newblock Human factors in software reliability engineering.
\newblock {\em Workshop on Applications of Human Error Research to Improve
  Software Engineering}, 2015.

\bibitem{spichkova2017human}
Maria Spichkova and Milan Simic.
\newblock Human-centred analysis of the dependencies within sets of proofs.
\newblock {\em Procedia computer science}, 112:2290--2298, 2017.

\bibitem{spichkova2016ahr}
Maria Spichkova and Anna Zamansky.
\newblock Ahr: Human-centred aspects of test design.
\newblock In {\em International Conference on Evaluation of Novel Approaches to
  Software Engineering}, pages 111--128. Springer, 2016.

\bibitem{spichkova2016human}
Maria Spichkova and Anna Zamansky.
\newblock A human-centred framework for supporting agile model-based testing.
\newblock In {\em CAiSE Forum}, pages 105--112, 2016.

\bibitem{tao2010automatic}
Qiuming Tao, Wei Wu, Chen Zhao, and Wuwei Shen.
\newblock An automatic testing approach for compiler based on metamorphic
  testing technique.
\newblock In {\em Asia Pacific Software Engineering Conference}, pages
  270--279. IEEE, 2010.

\bibitem{weyuker1982testing}
Elaine~J Weyuker.
\newblock On testing non-testable programs.
\newblock {\em The Computer Journal}, 25(4):465--470, 1982.

\bibitem{zamansky2016formal}
Anna Zamansky, Guillermo Rodriguez-Navas, Mark Adams, and Maria Spichkova.
\newblock Formal methods in collaborative projects.
\newblock In {\em Proceedings of the 11th International Conference on
  Evaluation of Novel Software Approaches to Software Engineering}, pages
  396--402, 2016.

\bibitem{zhu2015jfuzz}
Hong Zhu.
\newblock Jfuzz: A tool for automated java unit testing based on data mutation
  and metamorphic testing methods.
\newblock In {\em Trustworthy Systems and Their Applications}, pages 8--15.
  IEEE, 2015.

\end{thebibliography}

\end{document}